%% file: main.tex
\documentclass{article}
\usepackage{cite}
\usepackage{amsmath,amssymb,amsfonts}
\usepackage{algorithmic}
\usepackage{graphicx}
\usepackage{textcomp}
\usepackage{authblk}
\usepackage{geometry}
\def\BibTeX{{\rm B\kern-.05em{\sc i\kern-.025em b}\kern-.08em
    T\kern-.1667em\lower.7ex\hbox{E}\kern-.125emX}}

\begin{document}
\newcommand{\TiO}{TiO$_2$}
\newcommand{\SiO}{SiO$_2$}
\newcommand{\NanoRod}{Al-TiO$_2$-Al~}

\title{A new CMY camera technology using Al-\TiO-Al nanorod filter mosaic integrated on a CMOS image sensor}
\author[1]{X. He}
\author[1]{Y. Liu} 
\author[2]{P. Beckett}
\author[3]{H. Uddin}
\author[1]{A. Nirmalathas}
\author[1]{R. R. Unnithan}

\affil[1]{Department of Electrical and Electronic Engineering, The University of Melbourne, Melbourne, VIC, 3010, Australia.}
\affil[2]{School of Engineering, RMIT University, Melbourne, VIC, 3000, Australia.}
\affil[3]{Melbourne Centre for Nanofabrication, Australian National Fabrication Facility, Clayton, VIC, 3168, Australia.}



\maketitle
{Corresponding author: R. R. Unnithan (e-mail: r.ranjith@unimelb.edu.au)}

\input{Source/abstract.tex}
\input{Source/body.tex}

\input{Source/biblio.tex}
\end{document}

%% file: Source/abstract.tex
\begin{abstract}
A CMY colour camera differs from its RGB counterpart in that it employs a subtractive colour space of cyan, magenta and yellow. CMY cameras tend to performs better than RGB cameras in low light conditions due to their much higher transmittance. However, conventional CMY colour filter technology made of pigments and dyes are limited in performance for the next generation image sensors with submicron pixel sizes. These conventional filters are difficult to fabricate at nanoscale dimensions as they use their absorption properties to subtract colours. This paper presents a CMOS compatible nanoscale thick CMY colour mosaic made of Al-\TiO-Al nanorods forming an array 0.82 million colour pixels of 4.4 micron each, arranged in a CMYM pattern. The colour mosaic was then integrated onto a MT9P031 monochrome image sensor to make a CMY camera and the colour imaging demonstrated using a 12 colour Macbeth chart. The developed technology will have applications in astronomy, low exposure time imaging in biology and photography.

\textbf{Keywords: } CMY image sensor, Plasmonics, nanophotonics

\end{abstract}

%% file: Source/body.tex
\section{Introduction} \label{sec:introduction}
In a conventional CMOS based image sensor, colour imaging relies on the integration of filters on top of the photodetector array. These filters typically cover the three primary colours (bands): red (around 650 nm), green (550 nm) and blue (450 nm) (RGB), predominately in a Bayer pattern [1-9]. As the human eye is more sensitive to green light than either red or blue, the widely used Bayer filter mosaic is formed with twice as many green as red or blue filters. RGB colour space uses an additive colour mixing of red, green, and blue that combine to create a white output.

In contrast, CMY (cyan, magenta, and yellow) is a subtractive colour mixing scheme where colour filters are used to remove certain wavelengths of white light. For example, cyan is obtained when the red is subtracted from the image. Similarly, magenta and yellow are obtained by subtracting the green and blue respectively. In RGB space, a red filter transmits only about 1/3 of the visible light as the remaining light is absorbed in the filter (only the red light passes through the filter). In contrast, the corresponding cyan filter in CMY colour space transmits about 2/3 of the spectrum because only the red is subtracted, and the remaining is transmitted through the filter. In general, CMY colour filters pass approximately twice the spectral power as their corresponding RGB filters [11] and so are promising candidates for low-light imaging applications. Examples of such applications include astronomical imaging of nebula and the like that are characterized by dim objects against a dark background. Furthermore, astronomical images are often required to be captured with short exposure times, lest rotation effects blur the image (for example, images of Jupiter). These requirements are best served by high transmission filter schemes such as CMY.

Conventional CMY colour filters are based on the absorption properties of organic dye-based materials or pigments to subtract selected wavelengths from the incident light. However, as the pixel size in the sensor is reduced to submicron dimensions [8], conventional CMY colour filters start to suffer from colour cross talk as their performance deteriorates at nanoscale thicknesses [1-9]. Further, existing CMY technology must be fabricated in several steps, which presents severe challenges when trying to accomplish submicron-scale alignment. These issues demand the development of new CMY color filter technologies that can be fabricated easily at nanoscale thickness, using CMOS compatible materials, and that will support the creation of millions of filter pixels within a colour mosaic.

Colour filters based on plasmonic effects [1-9, 11-27] are suitable candidates due to their ability to be precisely tuned using nanoscale thick films. Colour filters based on localized surface plasmons are superior to those based on surface plasmons as the former are angle independent. As a result, the transmitted colours in LSP based filters are same for any angle of incidence, an important requirement for image sensor applications. Recently, gold (Au) nano-disk based CMY filters have been demonstrated with useful characteristics such as polarization independence and angle insensitivity [14]. However, Au based plasmonic filters are limited in their colour tuning capability, especially below 550 nm as Au does not readily support plasmonic resonance peaks below that wavelength [28]. 

CMY filters with high transmission coefficients have been reported based on silver (Ag) nano-slits [15], nano-disk [17] and Si meta-surfaces [18-19]. However, Ag oxidizes quickly in air and this degrades the optical properties, thereby mandating additional protective coatings that can further affect the filter characteristics. CMY filters operating in reflection mode based on surface plasmon polaritons (SPP) are demonstrated in [21-24], but the reflection mode is not suitable for image sensor applications. 

In this paper, we present a colour filter mosaic built from a hexagonal array of \NanoRod nanorods on a quartz substrate that is derived from a subtractive MDM (metal-dielectric-metal) nanohole array structure. The structure exhibits a high transmission efficiency and narrow bandwidth to produce superior colour separation. Colour tuning is achieved by varying the rod radius across the array, while keeping the base thickness constant, thereby forming a mosaic of CMY filters across the substrate. 
Further, we have applied this colour mosaic to develop a CMY camera. The 4mm x 4mm optical filter mosaic encompasses 0.82 million pixels, each pixel being 4.4 $\mu$m square and arranged in CMYM pattern. The filter has been integrated onto a monochrome image sensor (MT9P031) to create a CMY camera and its imaging capabilities demonstrated using a standard 12 colour Macbeth chart. To date, there has been no prior demonstration of a full camera system using a nanoscale thick CMY mosaic integrated onto an image sensor that illustrates its feasibility for imaging applications. This is the first such demonstration.

\section{Results}
 
\subsection{Filter Design and Fabrication}
\begin{figure}[t!]
    \centering
    \includegraphics[scale=0.4]{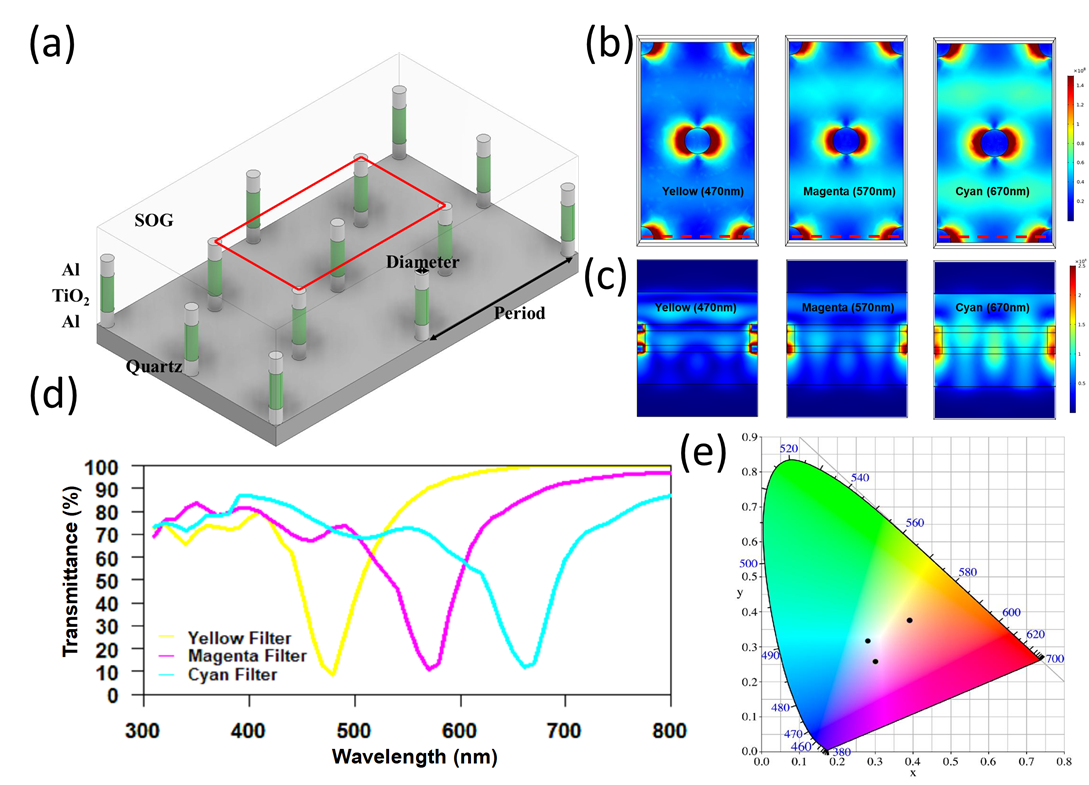}
    \caption{Simulation results of Al-\TiO-Al nanorod based CMY (cyan, magenta, yellow) filter mosaic: (a) Al-\TiO-Al nanorods in hexagonal array on a quartz substrate covered with spin-on-glass, (b) The normalized electric field at valley wavelength for the filters: yellow (470 nm), magenta (570 nm) and cyan (670 nm). Electric field at 470 nm, (c) The normalized electric field of cross section (red dotted line) at valley wavelength for the filters, (d) Numerically simulated transmission spectra of the CMY colour filters. The wavelength is swept from 300 nm to 800 nm (e) CIE chart of simulated CMY colours in the filter mosaic.\label{fig1}}
\end{figure}

Fig. \ref{fig1}a shows the proposed structure of the CMY colour filter, which is formed from Al-\TiO-Al nanorods fabricated on a quartz substrate and then embedded in a SOG (spin on glass [33]) matrix for refractive index matching. The optical characteristics of this structure can be considered to arise from a Fano resonance in which the top and bottom metal disk operate together as a coupled dipole, enhancing the magnetic field between them to produce both destructive and constructive (i.e., in-phase and anti-phase) interference modes. Choosing an appropriate radius and period for the structures results in a narrow reflection peak and a deep transmission valley that are sensitive to the symmetry of the dipole elements [32]. It has been shown in previous work related to nanodisks [27, 32] that these structures exhibit a broad (in-phase) resonance at smaller wavelengths and a narrow anti-phase resonance at larger wavelengths. This can be described in terms of a Q factor, which is proportional to the ratio of the resonant frequency to the FWHM. It is also clear that, while introducing asymmetries can strengthen the peaks (in particular, the anti-phase resonance), it introduces an undesirable sensitivity to the angle of incidence [32]. As a result, our work has kept the thickness and radius of both the top and bottom metal disks the same, which has the added advantage of simplifying its fabrication. 

The CMY filters were computationally investigated in 3D using finite element methods (FEM) implemented in COMSOL MULTIPHYSICS. The simulation model consists of a unit cell on a semi-infinite thick quartz substrate consisting of a single nanorod at the centre and one-quarter of a nanorod at each corner as shown in Fig. \ref{fig1}a (red rectangle). Each rod consists of an Al-\TiO-Al stack with thickness of 40 nm, 90 nm and 40 nm, respectively. The simulated unit cell on the semi-infinite glass substrate is covered with spin on glass. The 3D simulation geometry has been truncated using perfect matched layer (PML) and periodic boundary condition (PBC). The PML is applied on the top and at the bottom, PBCs are applied on the four sides of the cell as highlighted in the red block of Fig. \ref{fig1}a. For calculating the peak wavelengths, the refractive index of glass is set to 1.5, spin on glass 1.45 and the wavelength dependent refractive index of Al was taken from Rakic’s data [34]. The refractive index of \TiO~(i.e., 2.1) was obtained experimentally by measuring values from a \TiO~ film deposited with an E-beam evaporator. This value was also used in the simulation model. The transmission coefficient has been obtained from S-parameters using the port boundary conditions. 

The wavelength of light was swept from 300 nm to 800 nm to find the valley transmission [35], keeping the thickness values constant while varying the rod radius from 20 nm to 30 nm. The normalised electric field of each colour filter-yellow (470 nm), magenta (570 nm) and cyan (670 nm)—at their resonant wavelength (i.e., the valley wavelength) is shown in Figure 1b. The corresponding normalized electric field across each CMY pillar is shown Fig. \ref{fig1}c while Fig. \ref{fig1}d shows the simulated transmission spectrum for CMY filters. The layer parameters (thickness, period and rod diameter) are summarised in Table \ref{tab:I}. Fig. \ref{fig1}e plots the simulated CMY filter wavelengths on the CIE chromaticity chart, illustrating that these fall in an appropriate part of the colour space, thereby demonstrating their suitability for imaging applications.

\begin{figure}[t!]
    \centering
    \includegraphics[scale=0.4]{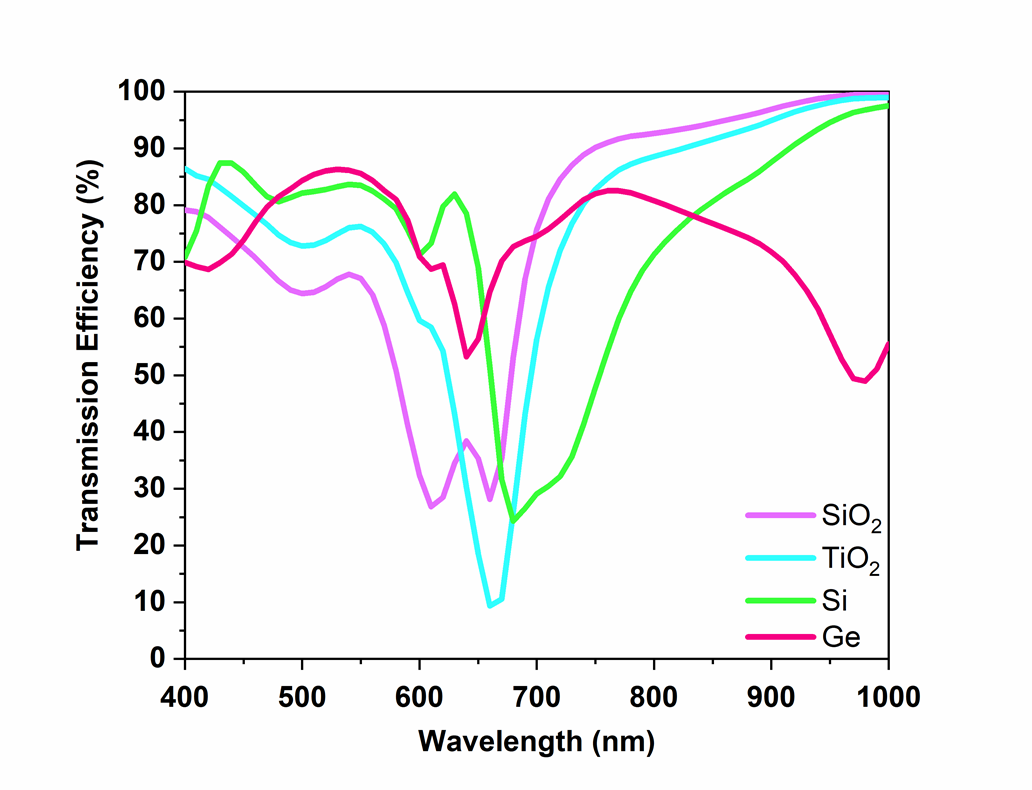}
    \caption{Comparison of transmission spectra from Al-X-Al, where X represents the middle dielectric material: SiO$_2$, \TiO, Si and Ge.\label{fig2}}
\end{figure}

\begin{table}[t!]
    \caption{Parameters of Al-\TiO-Al Nanorod (layer thickness, period, rod diameter)}
    \label{tab:I}
    \centering
    \begin{tabular}{l c c c}
    \hline \hline
         Parameters & Cyan & Magenta & Yellow \\
    \hline
       Top Al thickness (nm)&  40&40 &40 \\
       \TiO~thickness (nm) &90 &90&90\\
      Bottom Al thickness (nm) &  40&40 &40 \\
        Period (nm)& 500& 430& 350\\
        Diameter (nm) & 60 & 90&40 \\
    \hline
    \end{tabular}
\end{table}

\begin{figure}[t!]
    \centering
    \includegraphics[scale=0.3]{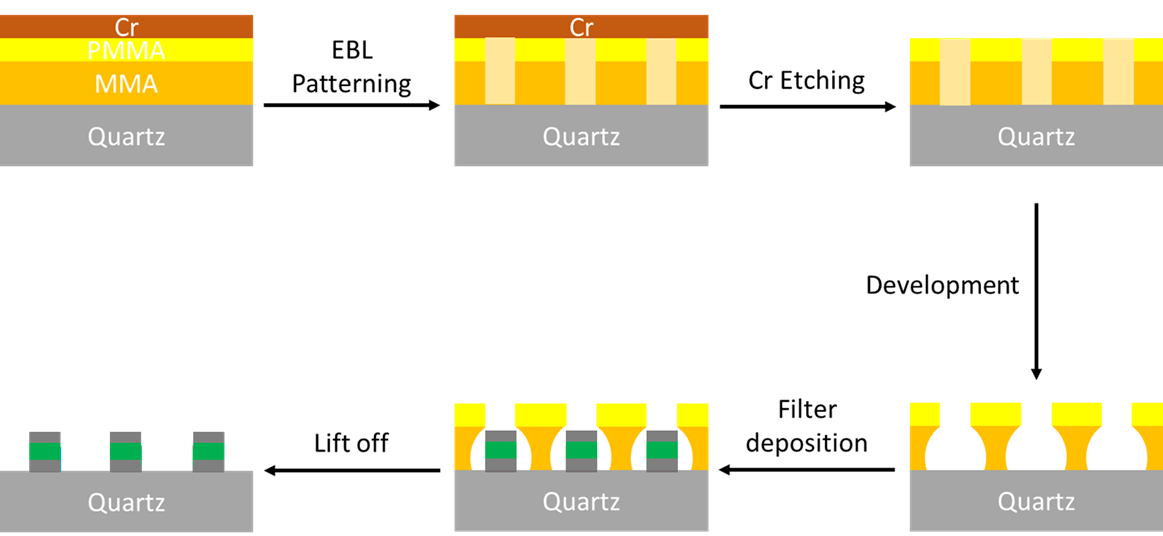}
    \caption{Strategy for fabricating the Al-\TiO-Al nanorod based CMY colour filter mosaic.\label{fig3}}
\end{figure}

As the refractive index of the middle (insulator) layer has substantial influence on the optical properties of the filter, we also investigated a number of alternative dielectric materials for this layer. In our simulation model, we applied SiO$_2$, \TiO, Si (assuming an ion assisted deposition with E-beam evaporator) and Ge with refractive index values of 1.45, 2.1, 2.8 and 4, respectively. These materials were used to simulate a cyan colour filter and the results are summarised in Figure 2. Note that these data are measured using ellipsometry (J.A. Woollam M-2000DI). As shown in Fig. \ref{fig2}, the transmission spectrum exhibits the largest transmission contrast and smallest linewidth (to produce a more vivid cyan colour) with \TiO~ compared to SiO$_2$, Si and Ge so this material was selected for the remainder of our experiments.

We fabricated a 4 mm x 4 mm mosaic onto a quartz substrate using standard nanofabrication techniques [36-39]. The mosaic comprises repeating CMYM units in both the horizontal and vertical directions to form a colour array. The fabrication strategy using a PMMA-MMA bilayer is described as follows (see Fig. \ref{fig3}).

Firstly, a 1 mm thick quartz substrate was cleaned with acetone under sonication for 5 minutes followed by IPA and DI water rinse. The substrate was then preheated at 80\textdegree C for 10 minutes and a thin EL9 (MMA) layer spin-coated onto the quartz at 3000 RPM for 1.5 minutes and baked at 180\textdegree for 15 minutes. A thin PMMA A2 layer was spin-coated onto the sample at 3000 RPM for 1.5 minutes and baked at 180\textdegree C for 5 minutes. To ameliorate problems with charging of the quartz substrate under Electron Beam Lithography (EBL), a 30 nm Cr layer was deposited on the sample by conformal sputtering to finish the sample preparation.

After patterning with EBL (Vistec EBPG5000plusES), the Cr layer was removed using 1 minute in a Cr etchant, stopped by 5\% H$_2$SO$_4$. The sample was then developed with diluted MIBK (MIBK:IPA=1:3) for 1 minute and stopped by IPA and DI water. It can be seen in Fig. \ref{fig3} that the MMA exhibits a bowl shape after development, which makes subsequent lift off easier due to the smaller contact area between the Al-\TiO-Al nanorod and the quartz. 

Finally, the Al-\TiO-Al nanorod was deposited by E-beam evaporator (Intlvac Nanochrome II). The deposition rate for the \TiO~ was 0.1 nm/s. After MMA/PMMA lift-off, the filter was completed by spin-coating with SOG (Desert NDG-2000) at 2000 RPM for 20 seconds and baking at 210\textdegree C for 15 minutes.

\begin{table}[t!]
    \caption{Optical Characteristics of the CMY Filter Mosaic}
    \label{tab:II}
    \setlength{\tabcolsep}{3pt}
    \centering
    \begin{tabular}{l c c c c c c}
    \hline \hline
 & \multicolumn{2}{c}{Cyan} & \multicolumn{2}{c}{Magenta} & \multicolumn{2}{c}{Yellow}\\
       Transmission Characteristic &Sim&Expt& Sim&Expt& Sim&Expt \\
        \hline
       Valley Wavelength (nm)  &660&670 &580&580&470&480\\
       Maximum Contrast (\%)&60&45&70&47&80&50\\
       Maximum Efficiency (\%)&90&90&100&92&100&90\\
       \hline
    \end{tabular}
\end{table}
 
\begin{figure}[t!]
    \centering
    \includegraphics[scale=0.3]{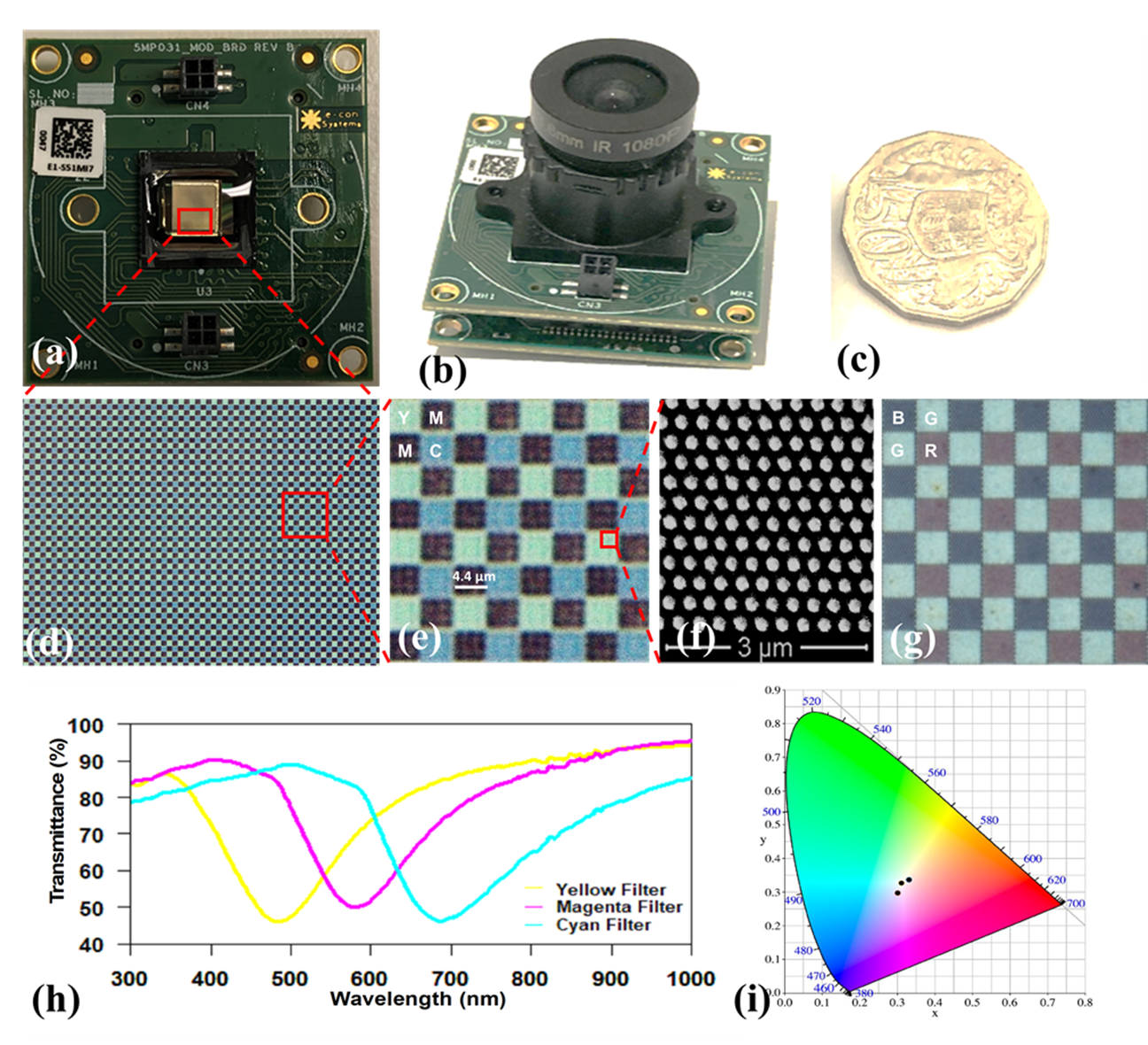}
    \caption{Integration of the CMY colour mosaic on the image sensor. (a) CMOS image sensor, MT9P031 integrated with 4 mm x 4 mm CMY colour mosaic using flip chip bonder for alignment; (b) CMY camera developed with optics ($f$ number 1.4) for imaging; (c) an Australian 50c coin as a reference to show the size of the CMY camera; (d) the CMY colour mosaic under optical microscope; (d) magnification $\times$20 (e) magnification $\times$40; (f) SEM image of the filter mosaic made of Al-\TiO-Al nanorods from top view (cyan is shown); (g) Reflected RGB colours from (e); (h) Experimental transmission spectra of the CMY color filters from the color mosaic; (i) CIE chromaticity chart of the CMY filters in the mosaic from experimental data.\label{fig4}}
\end{figure}
\begin{figure}[t!]
    \centering
    \includegraphics[scale=0.35]{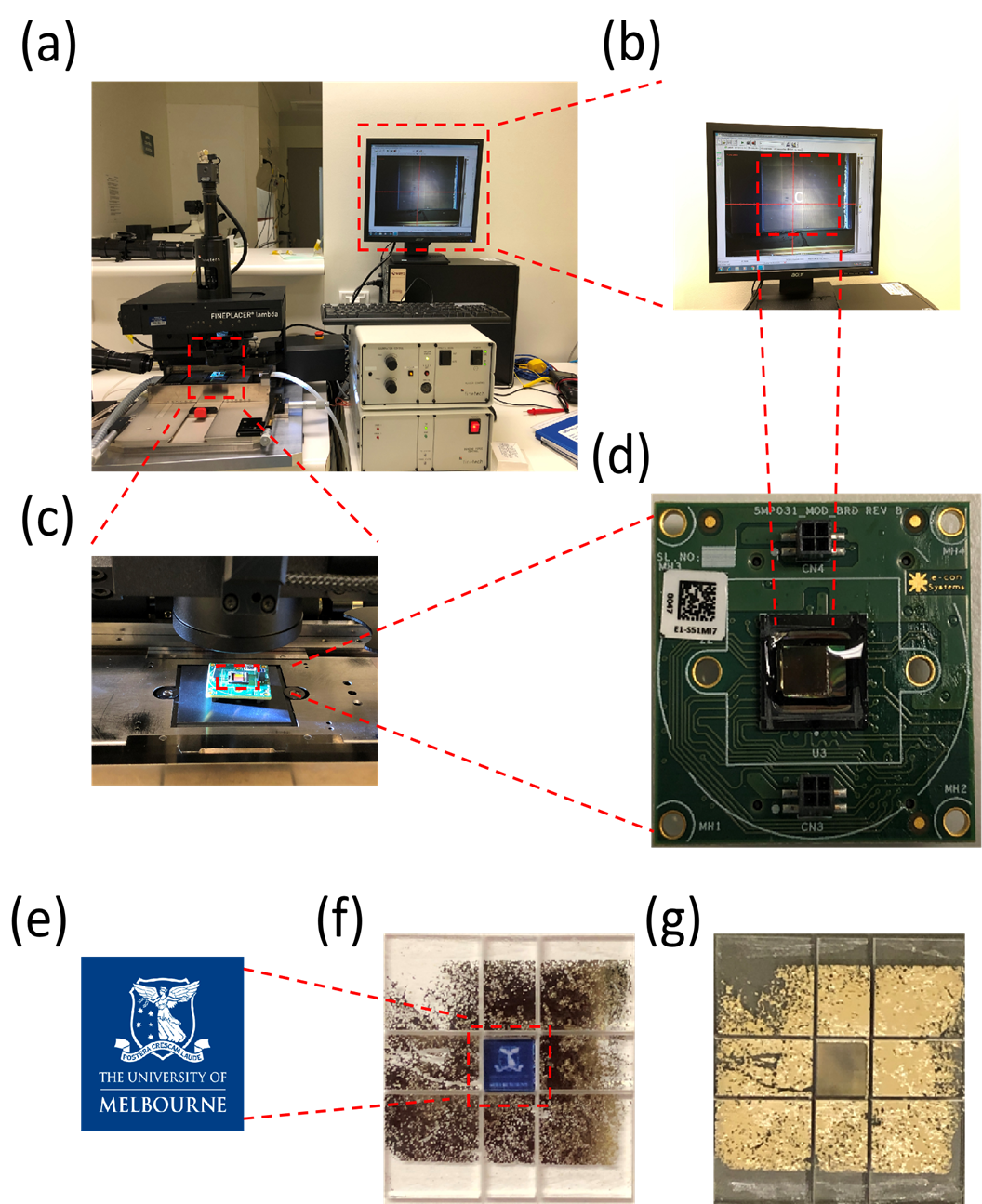}
     \caption{Integration of CMY filter mosaic on the CMOS image sensor MT9P031: (a) flip chip bonder, (b) display showing the alignment of filter colour pixels on the sensor pixels, (c) placement of the filter mosaic on the sensor (d) CMY filter aligned on the image sensor after being glued by PMMA powder diluted with anisole, (e) logo of the University of Melbourne (f) CMY filter with the logo behind it to show its high transmission efficiency, (g) CMY colour filter after being diced.\label{fig5}}
\end{figure}

Fig. \ref{fig4}d shows an image of the CMYM mosaic in transmission mode under an optical microscope (Olympus BX53M) with 20$\times$ magnification and an enlarged section of this is shown in Fig. \ref{fig4}e. A SEM image of part of a cyan pixel (top view) is shown in Fig. \ref{fig4}f. Note that, as the sample surface is non-conductive, 7nm of Cr was deposited prior to performing the SEM imaging, which altered the shape of the nanorods. Finally, it can be seen from the reflection mode image (Fig. \ref{fig4}g) that the reflected light from the sample (i.e., Fig. \ref{fig4}e) is RGB. 

The transmission spectrum of the fabricated mosaic was measured using a CRAIC spectrometer (Apollo Raman Microspectrometer). It can be seen from Fig. \ref{fig4}h that the maximum transmittance of the individual Cyan, Magenta and Yellow cells is around 90\%. The spectrum displays high transmission without any secondary resonances from UV to Near-IR wavelengths. Figure 4i positions these experimentally measured CMY data on the standard CIE chromaticity chart. The detailed results for the CMY filter are collated in Table II.

As is evident from Table \ref{tab:II}, generally good agreement was observed between the simulated and experimental data in the case of the valley wavelength and efficiency values. However, the peak transmission contrast is consistently smaller (by almost 40\% for yellow). This difference may be caused by the surface roughness of the deposited \TiO. It has been observed in prior work [27] that defects and surface roughness in the metal film can result in inhomogeneous broadening of the resonances without affecting the FWHM and can even eliminate the anti-phase mode entirely. In order to achieve a relatively uniform metal oxide layer, and therefore a high refractive index, the deposition rate of the E-beam evaporator needs to be set as high as possible. This resulted in a surface roughness of around 3-5 nm in our work (measured by Atomic Force Microscopy). In the absence of an additional planarization step, the top Al disk becomes similarly rough, potentially affecting its behavior. However, this reduction in contrast may be offset in part by the high transmission efficiency, which increases the received brightness. The contrast can also be corrected by subsequent image processing.

\subsection{Filter integration and Image Reconstruction}

The CMYM mosaic was integrated onto a commercial CMOS sensor using a flip-chip bonder (Fineplacer @Lambda) and raw images captured for further analysis as shown in Fig. \ref{fig5}. The CMOS image sensor used in this experiment was a MT9P031 monochrome array with a pixel size of 2.2 $\mu$m.
Fig. \ref{fig5}(a-d) shows the CMY mosaic aligned and glued onto the image sensor. The glue was made in-house using PMMA powder diluted with anisole solution and the filter was pressed onto the image sensor with just sufficient force to exclude most of the air between them. The high transmission efficiency of the CMY filter is demonstrated using the University logo on white paper which can be clearly seen through filter (Fig. \ref{fig5}(e-g)). The overall filter mosaic size is 4 mm x 4 mm and each colour cell covers a 2 x 2 patch to increase the light absorption and decrease the spatial crosstalk. Therefore, the pixel size in this mosaic is 4.4 $\mu$m square. The image sensor integrated with the mosaic is then fitted with optics ($f$ number = 1.4) for imaging as shown in Fig. \ref{fig4}b. The compact size of the CMY camera is shown by reference to an Australian 50c coin\footnote{A 12 sided coin measuring 31.65 millimetres between flat sides.} (Fig. \ref{fig4}c).

\begin{figure}[t!]
    \centering
    \includegraphics[scale=0.24]{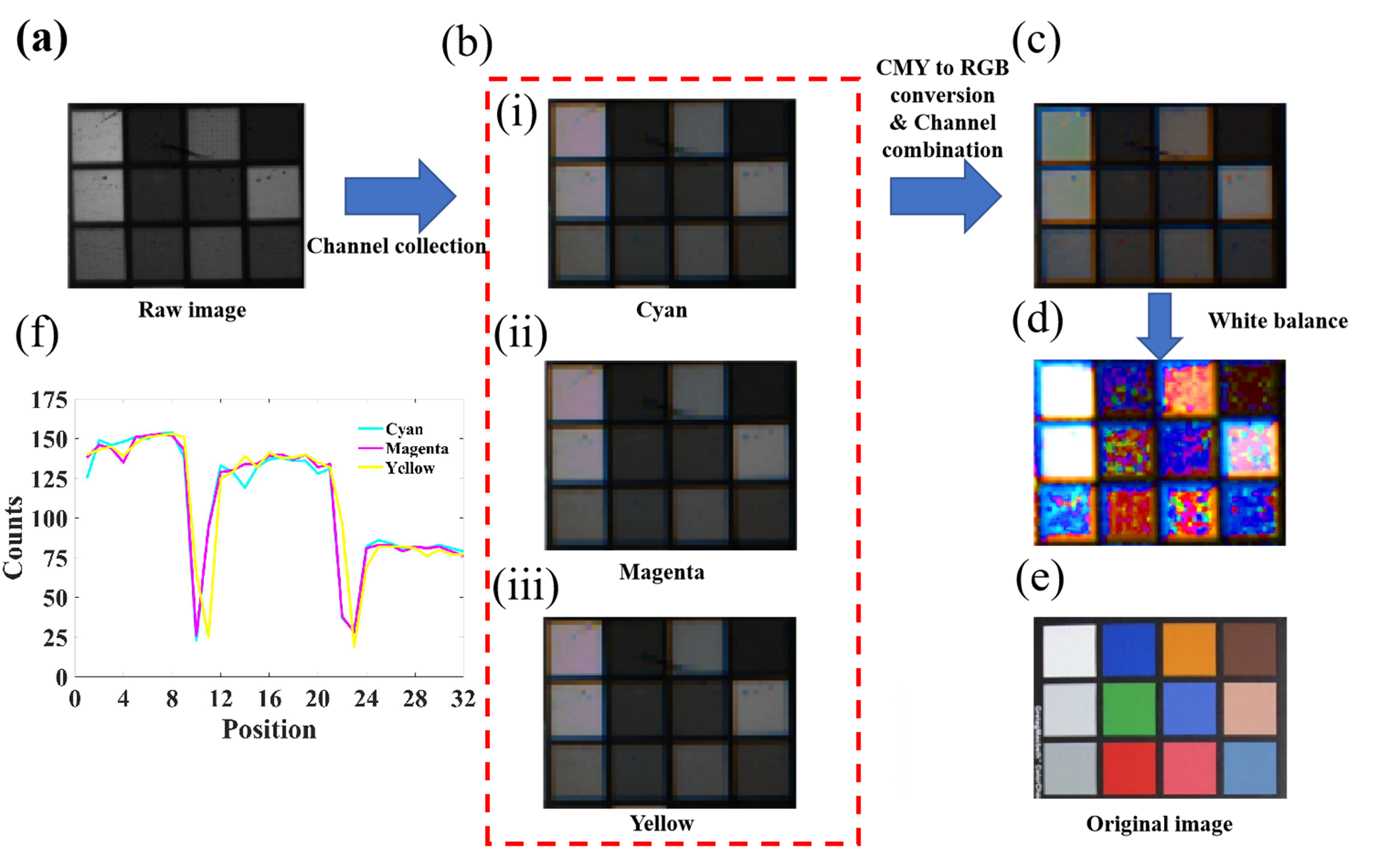}
    \caption{Demonstration of colour imaging by the CMY camera: (a) 12-bit raw image of the 12 colour Macbeth chart captured by CMY camera. (b) Three bands (cyan, magenta and yellow) extracted from the raw image. (c) The three bands (three channel combination) are recombined to get a CMY color image. (d) The CMY colour image is converted into RGB and applied color correction and white balance. (e) Standard image of the 12 colour Macbeth color chart. (f) The plot shows signals from three pixel numbers along the vertical red dotted line in the raw image.\label{fig6}}
\end{figure}

The CMY camera was then used in an example imaging application shown in Fig. \ref{fig6}. Firstly, a raw image of the standard 12-colour Macbeth chart (Figure 6e) was taken with the CMY camera (Fig. \ref{fig6}a). The recorded 12-bit raw image data was then transmitted to a laptop for processing using MATLAB. The raw data was first analyzed for saturation and signal intensity variation. Fig. \ref{fig6}f shows a plot of the signals from the 12-colour Macbeth chart pixels across the transect indicated by the red dashed line. The results show that the intensity variations are present in the raw image. Then, the individual CMY pixel data were grouped to form separate images for each of cyan, magenta and yellow pixels (Fig. \ref{fig6}b). Fig. \ref{fig6}c shows the reconstructed colour image. Next, a CMY to RGB conversion algorithm: R=Y+M-C; G=Y+C-M; B=C+Y-M [10], was applied to transform the CMY image to RGB for display. Lastly, colour correction and white balance were applied to recover the Macbeth chart as shown in Fig. \ref{fig6}e. These results demonstrate that colours can be retrieved from the 12-bit raw image data and illustrate correct operation of the CMY camera.

\section{Conclusion}

In conclusion, a CMY camera is demonstrated using a nanoscale CMY colour filter mosaic made of CMOS compatible \NanoRod nanorods with high transmission. The overall filter mosaic is 4 mm x 4 mm with each pixel occupying 4.4 $\mu$m square in a CMYM pattern and is integrated onto a MT9P031 CMOS image sensor using flip-chip bonding and an in-house prepared glue. The performance of the filter mosaic itself was first characterised and then the correct operation of the integrated CMY camera was illustrated using a 12 colour Macbeth chart as an object. The colours were retrieved using image processing algorithms. The use of more powerful image processing algorithms (as employed by commercial camera systems) will further improve the colour performance. This nanorod technology will overcome the limitations imposed by conventional colour filter technology for creating CMY colour image sensors and cameras with submicron pixel sizes. As a result, this technology is likely have applications ranging over astronomy, low exposure time imaging in biology and general photography.


\section*{Conflict of Interest}
The authors declare that they have no conflict of interest.

\section*{Acknowledgment}
This work was performed in part at the Melbourne Centre for Nanofabrication (MCN) in the Victorian Node of the Australian National Fabrication Facility (ANFF). The authors acknowledge financial support through ARC Discovery Project: DP170100363.

\section*{Author Contributions}

X. H. and R. R. U conceived the idea. X. H carried out theoretical design, fabricated the CMY mosaic filter and the optical measurements. X. H. Y. L performed the integration of the mosaic filter with image sensor. X. H. Y. L. and P. B. processed the image data and analyzed the results.  X. H and R.R.U wrote the paper. R. R. U, H.U and A.N supervised the project. All authors discussed the results and commented on the manuscript.